\documentclass[%
 reprint,
 amsmath,amssymb,
 aps,
]{revtex4-1}

\usepackage{graphicx}
\usepackage{dcolumn}
\usepackage{bm}

\begin{document}

\preprint{APS/123-QED}


\title{Fluctuation of the Hubble Parameter}

\author{Xi-Bin Li$^1$}

\author{Hao-Feng Qin$^1$}

\author{Zhi-Song Zhang$^2$}

\author{Tong-Jie Zhang$^{1,3,4}$}
\email{tjzhang@bnu.edu.cn}

\affiliation{$^1$Department of Astronomy, Beijing Normal University, Beijing, 100875, China}
\affiliation{$^2$Department of Aerospace Engineering, School of Astronautics, Harbin Institute of Technology (HIT), Harbin Heilongjiang, 150001, China}
\affiliation{$^3$Departments of Physics and Astronomy, University of California, Berkeley, CA 94720, USA\\
$^4$Lawrence Berkeley National Laboratory, 1 Cyclotron Road, Berkeley, CA 94720, USA}

\date{\today}

\begin{abstract}
We study the Hubble parameter $H(z)$ in perturbed Friedmann universe and obtain an expression of the perturbed Hubble parameter $H(z,\textbf{n})$. We derive the Hubble parameter power spectrum by using the initial spectrum during inflation and the Bardeen transfer function. We obtain a semi-analytical expression in the case of cold dark matter (CDM) universe. Similar with luminosity distance, the Hubble parameter spectrum is suggested to be an useful observational tool to determine some cosmological parameters. In addition, we show that the Hubble parameter power spectrum could be used to check whether the expansion is accelerated by the second order small scale fluctuation.\\
\begin{description}
\item[PACS numbers]
98.80.-k, 98.62.Py, 98.80.Es, 95.36.+x
\end{description}
\end{abstract}

\pacs{Valid PACS appear here}
\maketitle


\section{\label{introduction}Introduction}

The Hubble parameter $H(z)$ as a direct quantitative measure of cosmic expansion rate has been studied in a large number of cosmological observations, like mapping the cosmic microwave background (CMB) anisotropy \cite{1,2}, measurement on the baryon acoustic oscillation (BAO) \cite{3,4} and some other ways \cite{5,6,7,8}. Recently, the Hubble parameter is widely used to constrain the cosmological parameters, such as dark energy, dark matter, and dark age \cite{9,10}.

During recent years, measurement on distant Ia supernova \cite{11} and anisotropy of the CMB combined with the large scale structure \cite{12} reveal that our universe is on an accelerated expansion. Since such acceleration was discovered, dark energy has led to a heated debate among researchers. Some believe that dark energy is responsible for the accelerated expansion \cite{13,14,15}. On the other hand, others reckon that second order perturbation might cause the acceleration instead of the cosmological constant \cite{16,17}. This claim seems a little surprising, since it requires big perturbation led by backreaction, contrary to the observation in the CMB. Another problem being under discussed is that whether dark energy is inhomogeneous \cite{18,19} and whether such fluctuation would lead to the acceleration \cite{20,21}.

To solve the problems above, more accurate observation and further theoretical study are needed. Observation on fluctuation of observational cosmological parameters has become a valuable practice. For example, fluctuation of the CMB is an excellent method to constrain some cosmological parameters \cite{insert} and global anisotropy of the CMB may be connected with the anisotropy of dark energy \cite{22,23}. Recently, the luminosity distance power spectrum has been used to test the fluctuation of dark energy \cite{24,25}. Camille Bonvin et al. \cite{26} calculated the luminosity distance power spectrum in perturbed Friedmann universe and used it to test whether second order perturbation leads to the accelerated expansion. For the study on fluctuation of expansion rate, although some work have predicted the fluctuation of the Hubble parameter \cite{27,28}, deep researches have not been in process yet.

In this paper, we attempt to illustrate the existence of the fluctuation of expansion rate. We work in the frame of a perturbed Friedmann universe and derive the Hubble parameter power spectrum within linear perturbation theory. The main point of our work is that the expansion rate, instead of other averaged cosmological parameter, can directly determine the acceleration parameter. The magnitude of the Hubble parameter power spectrum can determine the role of second order perturbation in the expansion universe and further determine whether dark energy dominates in the universe. What's more, we propose this power spectrum as a new observational method that may help to study dark energy and can also be used to constrain some cosmological parameters.

We work in a simple condition without the cosmic constant, and we reckon it is valuable. First, $\textrm{CDM}$ allows the analytical expression of the transfer function, the growth function as well as the conformal Hubble parameter. Then, it is enough to show the existence of such fluctuation. Besides, this work also sheds some light into the study on the fluctuation of the Hubble parameter in $\Lambda\textrm{CDM}$. Although we focus on the $\textrm{CDM}$ universe, it will not fade the value of the new method.

This paper is organized as follows: In Sec. \ref{sec2} we give a general expression of the Hubble parameter in a perturbed CDM universe. In Sec. \ref{sec3} we calculate the power spectrum by using the initial scale-invariant spectrum and Bardeen transfer function. Next, by employing some approximations, we obtain the analytical expression of Hubble parameter power spectrum in Sec. \ref{sec4}, and we also plot spectrums at different redshifts. Finally, we conclude in Sec. \ref{sec5} and discuss the observational issue.

\section{\label{sec2}Hubble parameter in perturbed universe}
The metric for a space-time with homogeneous and isotropic sections is the maximally-symmetric Robertson-Walker(RW) metric \cite{9}, which can be written in the form
\begin{eqnarray}
ds^2=-dt^2+a^2(t)[\frac{dr^2}{1-kr^2}+r^2d\theta^2+r^2sin^2\theta d\phi^2],
\end{eqnarray}
where $a(t)$ is the expansion scale factor of the universe and $k$ is the curvature of the universe. Using the Einstein equation, we obtain
\begin{eqnarray}
    H^2&=&(\frac{da/dt}{a})^2=\frac{8\pi G}{3}(\rho_M+\rho_R)+\frac{\Lambda c^2}{3}-\frac{kc^2}{a^2} \nonumber \\
    &=&H^2_0[\Omega_M(1+z)^3+\Omega_R(1+z)^4+\Omega_\Lambda+\Omega_k(1+z)^2],  \nonumber \\  \label{eq2}
\end{eqnarray}
where $H\equiv({da/dt})/{a}$ is the Hubble parameter and $H_0$ is called the Hubble constant which represents the value of the Hubble parameter at present time. $\Omega_M$, $\Omega_R$, $\Omega_\Lambda$ and $\Omega_k$ are cosmological density parameter of matter, radiation, dark energy and curvature respectively at present epoch. In this paper, we consider the cold dark matter (CDM) universe. So except $\Omega_M=1$, other parameters vanish.

Turn to the perturbed model. In the flat RW background, metric of scalar perturbation with Newtonian (longitudinal) gauge is
\begin{eqnarray}
    g_{\mu\nu} dx^\mu dx^\nu=-(1+2\Psi)dt^2+a^2(t)(1-2\Phi)\delta_{ij} dx^i dx^j .\nonumber \\
\end{eqnarray}
Metric perturbation $\Phi$ and $\Psi$ are equal for perfect fluids. In geometry dynamics, the Hubble parameter is defined as one third of covariant derivative of time-like unit vector $n^\mu$, i.e. $H=n^\mu_{;\mu}/3$ \cite{Bojowald}, where $n^\mu$ as a function of time and position is also orthogonal to spacial hypersurface. In fact, $n^\mu$ coincides with the velocity field of matter in unperturbed background \cite{perturbation}. This invokes us to study the fluctuation of Hubble parameter through the 4-velocity field. In practice, the Hubble parameter is measured via the luminosity distance to standard candles. In Appendix B, we also derive such fluctuation via derivative of the perturbed luminosity distance.

More directly to understand the perturbed Hubble parameter, in unperturbed universe the Hubble parameter $H(z)$ is independent of position. Once perturbation is considered, gravitational potential $\Psi$ makes the local scale factor $a(t)$ amplified or compressed and further leads to the change of local expansion rate. It is easy to imagine that our universe in overdensity regime expands not so fast as it does in underdensity regime. This phenomenon is mainly determined by the gravitational potential $\Psi$.

First we introduce the proper time $d\tau=\sqrt{-g_{00}}dt$. And consider $a(t)(1-2\Psi)^{1/2}$ as the modified scale factor. Simulating the definition of the Hubble parameter in unperturbed universe, we define the Hubble parameter in perturbed universe as
\begin{eqnarray}
    H(\textbf{n},z)=\frac{\frac{d}{d\tau}\left[ a(t)(1-2\Psi)^{1/2}\right]} {a(t)(1-2\Psi)^{1/2}}.
\end{eqnarray}
This seems a little strange.

Then we get the Hubble parameter from velocity field. The relationship between the comoving distance $\textbf{x}$ and physical distance $\textbf{X}$ at arbitrary cosmic time $t$ is
\begin{eqnarray}\textbf{X}(t,\textbf{x})=a(t)(1-\Psi)\textbf{x}. \label{eq01}\end{eqnarray}
Then taking derivative of Eq.~(\ref{eq01}) with respect to proper time $\tau$, we get the velocity
\begin{eqnarray}
    \textbf{v}(t,\textbf{x})=\frac{\frac{d}{d\tau}\left[ a(t)(1-\Psi)\right]} {a(t)(1-\Psi)}\textbf{X}(t,\textbf{x})+\textbf{v}_p. \label{insert5}
\end{eqnarray}
The first term on the right hand is the Hubble law and the second term $\textbf{v}_p$ is the peculiar velocity which is given by \cite{insert1}
\begin{eqnarray}
    \textbf{v}_p=\frac{1}{4\pi G(\rho+p)}(\frac{\dot{a}}{a}\partial_i\Psi+\partial_i\dot{\Psi}). \label{velocity}
\end{eqnarray}
Normally, $\textbf{v}_p$ is absorbed into the fluctuation of redshift \cite{insert3,26}.

Comparing Eq.~(\ref{insert5}) with the Hubble law, we get the Hubble parameter in perturbed CDM universe
\begin{eqnarray}
    H(\textbf{n},z)&\equiv &\frac{\frac{d}{d\tau}\left[ a(t)(1-2\Psi)^{1/2}\right]} {a(t)(1-2\Psi)^{1/2}} \nonumber \\
    &=&\overline{H}(\tilde{z})-\overline{H}(\tilde{z})\Psi+a^{-1}(t)\frac{d}{d \eta}\Psi.  \label{insert7}
\end{eqnarray}
Here $\overline{H}(\tilde{z})$ is the unperturbed Hubble parameter, and $\eta$ is the conformal time with $d\eta={dt}/{a(t)}$. It is convenient to employ the conformal position $\textbf{x}=\textbf{x}_\textbf{0}+\textbf{n}(\eta_0-\eta)$ when calculating spectrums, where $\textbf{x}_0$ is our position and $\textbf{n}$ is the unit vector along direction of the galaxy we observe. So the derivative of conformal time $\eta$ becomes $d/d\eta=\partial/\partial\eta-\textbf{n}\cdot\nabla$.

However, the quantity $\tilde{z}=1/a(\eta)-1$ is not directly measurable. And what can be measured instead is the quantity $z=\tilde{z}+\delta z$. Now
\begin{eqnarray}
    H(\textbf{n},\eta(\tilde{z}))=H(\textbf{n},\tilde{z})=H(\textbf{n},z)-\frac{d}{d\tilde{z}}H(\textbf{n},z)\delta z.
\end{eqnarray}
Moreover
\begin{eqnarray}
    \frac{d}{d\tilde{z}}H(\textbf{n},z)=\frac{3}{2}(1+z)^{-1}H(\textbf{n},z)+\textrm{first order}
\end{eqnarray}
for CDM and
\begin{eqnarray}
    \delta z=(1+z)\Big[& &\Psi(\eta_0)-\Psi(\eta) \nonumber\\
    & &+2\int^{\eta}_{\eta_0}{d\eta \dot{\Psi}}+(\textbf{v}_0-\textbf{v})\cdot\textbf{n} \Big], \label{ptbz}
\end{eqnarray}
where $\cdot$ donates the derivative with respect to conformal time ${\partial}/{\partial \eta}$. The first term $\Psi(\eta_0)-\Psi(\eta)$ on the right hand of Eq.~(\ref{ptbz}) is the modification on scale factor, the second on optics travelling through the gravity potential while the third on Doppler effect \cite{26}. $\Psi(\eta_0)$ and $\textbf{v}_0\cdot\textbf{n}$ mainly contribute to the dipole of spectrum which could be detected by measuring the dipole of the CMB \cite{insert2}, so we neglect these two terms. Inserting the above into Eq.~(\ref{insert7}), the Hubble parameter reads
\begin{eqnarray}
    H(\textbf{n},z)&=&\overline{H}(z)+\frac{1}{2}\overline{H}(z)\Psi-3\overline{H}(z)\int^{\eta}_{\eta_0}{d\eta \dot{\Psi}}\nonumber\\
    & &+\frac{3}{2}\overline{H}(z)\textbf{v}\cdot\textbf{n}+\frac{1}{a(t)}\frac{\partial}{\partial \eta}\Psi-\frac{1}{a(t)}\textbf{n}\cdot\nabla \Psi. \nonumber\\ \label{eq51}
\end{eqnarray}
Finally fluctuation of the Hubble parameter can be written as
\begin{eqnarray}
   \delta_H(\textbf{n},z)&\equiv& \frac{H(\textbf{n},z)-\overline{H}(z)}{\overline{H}(z)} \nonumber \\
    &=&\frac{1}{2}\Psi-3\int^{\eta}_{\eta_0}{d\eta \dot{\Psi}}+\frac{3}{2}\textbf{v}\cdot\textbf{n}\nonumber\\
    & &\quad+\frac{1}{\mathcal{H}(\eta)}\dot{\Psi}-\frac{1}{\mathcal{H}(\eta)}\textbf{n}\cdot\nabla \Psi  , \label{eq5}
\end{eqnarray}
where $\mathcal{H}$ is the conformal Hubble parameter with $\mathcal{H}\equiv{\dot{a}}/{a}={da}/{dt}=a(\eta) H(\eta)={2}/{\eta}$.

\section{\label{sec3}Hubble parameter power spectrum}
Power spectrum $C_\ell(z,z')$ is defined as
\begin{eqnarray}\delta_H(\textbf{n},z)=\sum_{\ell m}{a_{\ell m}(z)Y_{\ell m}(\textbf{n})},\end{eqnarray}
\begin{eqnarray}C_\ell(z,z')=\langle a_{\ell m}(z)a^*_{\ell m}(z')\rangle,\end{eqnarray}
where $\langle\cdot\rangle$ means the statistical average over whole space. The Fourier transfer of gravitational potential is
\begin{eqnarray}\Psi(\textbf{x})=\frac{1}{(2\pi)^3}\int{d^3ke^{i\textbf{k}\cdot\textbf{x}}\Psi(\textbf{k})},\end{eqnarray}
and
\begin{eqnarray}\Psi(\textbf{k})=\int{d^3xe^{-i\textbf{k}\cdot\textbf{x}}\Psi(\textbf{x})}. \label{eq8}\end{eqnarray}
The evolution of gravitational potential $\Psi$ can be separated into several parts \cite{29,wlde}
\begin{eqnarray}\Psi(\eta,\textbf{k})=T(\textbf{k})\frac{D(a)}{a}\Psi(\textbf{k}). \label{eq9}\end{eqnarray}
Here $T(k)$ is the Bardeen transfer function, and $D(a)$ is the growth function. $T(k)$ is normalized such that $T(k)\rightarrow 1$ for $k\rightarrow 0$. The growth function satisfies that $D(a)\rightarrow 1$ for $a\rightarrow 1$. And $D(a)=a$ in $\textrm{CDM}$ at late epoch($ z\leq 10 $) so $D(a)/a=1$. $\Psi(k)$ is the potential during inflation at horizon-cross scale. Besides, we consider only linear evolution.

Two point correlation function is
\begin{eqnarray}\zeta(|\textbf{x}-\textbf{y}|)=\langle\Psi(\textbf{x})\Psi(\textbf{y})\rangle\end{eqnarray}
Using Eq.~(\ref{eq8}) and ~(\ref{eq9}), we have
\begin{eqnarray}
    &&\langle\Psi(\eta,\textbf{k})\Psi^*(\eta',\textbf{k}')\rangle\nonumber \\
    &=&T(k)T(k')\int{d^3xd^3y\zeta(|\textbf{x}-\textbf{y}|)e^{-i\textbf{k}\cdot\textbf{x}}e^{i\textbf{k}'\cdot\textbf{y}}}\nonumber \\
    &=&(2\pi)^3T(k)T(k')k^{-3}P_\Psi(k)\delta^3(\textbf{k}-\textbf{k}'),
\end{eqnarray}
where $k^{-3}P_\Psi(k)$ is the initial spectrum of $\Psi(k)$
$$k^{-3}P_\Psi(k)=\int{d^3x\zeta(|\textbf{x}|)e^{-i\textbf{k}\cdot\textbf{x}}}.$$ So
\begin{eqnarray}
    &&k^3\langle\Psi(\eta,\textbf{k})\Psi^*(\eta',\textbf{k}')\rangle\nonumber \\
    &=&(2\pi)^3T(k)T(k')P_\Psi(k)\delta^3(\textbf{k}-\textbf{k}').
\end{eqnarray}
Standard inflation model predicts $P_\Psi(k)\simeq A(k/aH)^{n_s-1}$. Recent Planck data and other observations on anisotropy of CMB \cite{31,1} show that $n_s\simeq1$ and $A\sim10^{-10}$. In order to determine the correlation function, we introduce the conformal position $\textbf{x}=\textbf{x}_0+\textbf{n}(\eta_0-\eta)$ and $\textbf{x}'=\textbf{x}_0+\textbf{n}'(\eta_0-\eta')$. In this case, we have

\begin{eqnarray}& &\langle\Psi(\eta,\textbf{x})\Psi(\eta',\textbf{x}')\rangle\nonumber \\
    &=&\frac{1}{(2\pi)^6}\int{d^3kd^3k'T(k)T(k')\langle\Psi(\textbf{k})\Psi^*(\textbf{k}')\rangle}\nonumber \\
    & &\quad\quad\quad\quad\times e^{i\textbf{k}\cdot\textbf{n}(\eta_0-\eta)}e^{-i\textbf{k}'\cdot\textbf{n}'(\eta_0-\eta')}
\end{eqnarray}
Using the identity Eq.~(\ref{eq:B1}) we have
\begin{widetext}
\begin{eqnarray}
\langle\Psi(\eta,\textbf{x})\Psi(\eta',\textbf{x}')\rangle
&=&\frac{1}{(2\pi)^3}\sum_{\ell\ell'mm'}(4\pi)^2i^{\ell'-\ell}\int{\frac{dk}{k}T^2(k)P_\Psi(k)
 j_\ell(k(\eta_0-\eta))j_\ell(k(\eta_0-\eta'))}\nonumber \\
 & &\times Y^*_{\ell m}(\textbf{n})Y_{\ell' m'}(\textbf{n}')\int{d\Omega_\textbf{k}Y_{\ell m}(\hat{\textbf{k}})Y^*_{\ell' m'}(\hat{\textbf{k}})}\nonumber   \end{eqnarray}
\begin{eqnarray}
 &=&\frac{2}{\pi}\sum_{\ell m}Y^*_{\ell m}(\textbf{n})Y_{\ell m}(\textbf{n}')\int{\frac{dk}{k}T^2(k)P_\Psi(k)
 j_\ell(k(\eta_0-\eta))j_\ell(k(\eta_0-\eta'))}\nonumber \\
 &=&\frac{1}{2\pi^2}\sum_{\ell}(2\ell+1)P_\ell(cos\theta)\int{\frac{dk}{k}T^2(k)P_\Psi(k)j_\ell(k(\eta_0-\eta))j_\ell(k(\eta_0-\eta'))}\nonumber\\
 &=&\sum_{\ell}\frac{2\ell+1}{4\pi}C_\ell^{(\Psi)}(z,z')P_\ell(\textbf{n}\cdot\textbf{n}').
\end{eqnarray}
where $\hat{\textbf{k}}$ is the unit vector along wave number $\textbf{k}$, and in last equal equality Eq.~(\ref{eq:B8}) has been used.

According to Eq.~(\ref{eq:B1}), taking derivative with respect to $\eta_0-\eta$ of the complex exponential function raises the term $i\textbf{n}\cdot\textbf{k}$, i.e. $\textbf{n}\cdot\nabla\Psi(\eta,\textbf{x})=i\textbf{n}\cdot\textbf{k}\Psi(\eta,\textbf{x})$. So we just need to replace $j_\ell(k(\eta_0-\eta))$ by $kj'_\ell(k(\eta_0-\eta))$ in the terms containing $\textbf{n}\cdot\nabla\Psi$. From the expression of $\delta_H(\textbf{n},z)$, we separate $\langle\delta_H(\textbf{n},z)\delta_H(\textbf{n}',z')\rangle$ into several parts as follow
\begin{eqnarray}
\langle\Psi(\eta,\textbf{x})\Psi(\eta',\textbf{x}')\rangle&=&\sum_\ell
    \frac{2\ell+1}{4\pi}C_\ell^{(\Psi\Psi')}P_\ell(\textbf{n}\cdot\textbf{n}') \qquad \textrm{with} \nonumber \\
    C_\ell^{(\Psi\Psi')}&=&\frac{2}{\pi}\int{\frac{dk}{k}T^2(k)P_\Psi(k)j_\ell(k(\eta_0-\eta))j_\ell(k(\eta_0-\eta'))},\\
\langle\dot{\Psi}(\eta,\textbf{x})\Psi(\eta',\textbf{x}')\rangle&=& \sum_\ell
    \frac{2\ell+1}{4\pi}C_\ell^{(\eta\Psi')}P_\ell(\textbf{n}\cdot\textbf{n}') \qquad \textrm{with} \nonumber \\
    C_\ell^{(\eta\Psi')}&=&-\frac{2}{\pi}\frac{1}{\mathcal{H}(\eta)}\frac{\partial}{\partial \eta}\int{\frac{dk}{k}T^2(k)P_\Psi(k)
    j_\ell(k(\eta_0-\eta))j_\ell(k(\eta_0-\eta'))},\\
\langle\dot{\Psi}(\eta,\textbf{x})\dot{\Psi}(\eta',\textbf{x}')\rangle&=&\sum_\ell
    \frac{2\ell+1}{4\pi}C_\ell^{(\eta\eta')}P_\ell(\textbf{n}\cdot\textbf{n}') \qquad \textrm{with} \nonumber \\
    C_\ell^{(\eta\eta')}&=&\frac{2}{\pi}\frac{1}{\mathcal{H}(\eta)\mathcal{H}(\eta')}\frac{\partial ^2}{\partial \eta\partial \eta'}
    \int{\frac{dk}{k}T^2(k)P_\Psi(k)j_\ell(k(\eta_0-\eta))j_\ell(k(\eta_0-\eta'))},\\
\langle\textbf{n}\cdot\nabla\Psi(\eta,\textbf{x})\Psi(\eta',\textbf{x}')\rangle&=&\sum_\ell
    \frac{2\ell+1}{4\pi}C_\ell^{(n\Psi')}P_\ell(\textbf{n}\cdot\textbf{n}') \qquad \textrm{with} \nonumber \\
    C_\ell^{(n\Psi')}&=&\frac{2}{\pi}\int{dkT^2(k)P_\Psi(k)j'_\ell(k(\eta_0-\eta))j_\ell(k(\eta_0-\eta'))},\label{spectrum4}\\
\langle\textbf{n}\cdot\nabla\Psi(\eta,\textbf{x})\dot{\Psi}(\eta',\textbf{x}')\rangle&=&\sum_\ell
    \frac{2\ell+1}{4\pi}C_\ell^{(n\eta')}P_\ell(\textbf{n}\cdot\textbf{n}') \qquad \textrm{with} \nonumber \\
    C_\ell^{(n\eta')}&=&-\frac{2}{\pi}\frac{1}{\mathcal{H}(\eta')}\frac{\partial}{\partial \eta'}\int{dkT^2(k)P_\Psi(k)j'_\ell(k(\eta_0-\eta))j_\ell(k(\eta_0-\eta'))},\label{eqC2}\\
\langle\textbf{n}\cdot\nabla\Psi(\eta,\textbf{x})\textbf{n}\cdot\nabla\Psi(\eta',\textbf{x}')\rangle&=&\sum_\ell \frac{2\ell+1}{4\pi}C_\ell^{(nn')}P_\ell(\textbf{n}\cdot\textbf{n}') \qquad \textrm{with} \nonumber \\
    C_\ell^{(nn')}&=&\frac{2}{\pi}\int{dkkT^2(k)P_\Psi(k)j'_\ell(k(\eta_0-\eta))j'_\ell(k(\eta_0-\eta'))}.\label{eqC3}
\end{eqnarray}
\\
\\
With the expressions above, we can rewrite correlation function of Hubble parameter as
\begin{eqnarray}\langle\delta_H(\textbf{n},z)\delta_H(\textbf{n}',z')\rangle=\sum_\ell \frac{2\ell+1}{4\pi}P_\ell(cos\theta)(C_\ell^{(1)}+C_\ell^{(2)}+C_\ell^{(3)}).\end{eqnarray}
\end{widetext}
where $C_\ell^{(i)}$($i$=1, 2 and 3) represents the terms containing different number of Doppler factor $\textbf{n}\cdot\nabla\Psi(\eta,\textbf{x})$. $C_\ell^{(1)}$ contains no Doppler factor and represents the integral of $\Psi$ and its derivative of $\eta$.  $C_\ell^{(2)}$ contains one Doppler factor and represents the integral of $\textbf{n}\cdot\nabla\Psi$, $\Psi$ and $\dot{\Psi}$. $C_\ell^{(3)}$ contains two Doppler factors and represent the integral of only $\textbf{n}\cdot\nabla\Psi$. Preliminary result shows that $C_\ell^{(3)}$ dominates in $C_\ell^{(tot)}$ and we name it as Doppler term. But we will also see that $C_\ell^{(3)}$ is not the only term containing the Doppler term. The detailed expression of $C_\ell^{(i)}$'s are given in Appendix A.

One could find that $\int^{\eta}_{\eta_0}{d\eta \dot{\Psi}}$ and $\textbf{v}\cdot\textbf{n}$ are not taken into consideration, because $\langle \Psi\cdot\int^{\eta'}_{\eta_0}{d\eta \dot{\Psi}}\rangle \sim \langle \partial_\eta\Psi \cdot \int^{\eta'}_{\eta_0}{d\eta \dot{\Psi}}\rangle \sim 10^{-11} \ll \langle\textbf{n}\cdot\nabla\Psi(\eta,\textbf{x})\textbf{n}\cdot\nabla\Psi(\eta',\textbf{x}')\rangle $\cite{26} and $\textbf{v}\cdot\textbf{n}$ can be absorbed into the power spectrum of the luminosity distance which mainly contributes to dipole. So these two terms above are neglected in the Hubble parameter spectrums.

\section{\label{sec4}Approximation and Results}
Now we calculate the Hubble parameter power spectrum in CDM universe. In this model $\Omega_r$, $\Omega_k$, $\Omega_\Lambda$ in Eq.~(\ref{eq2}) vanish, except that $\Omega_m=1$. Standard inflationary scenarios give the scale-invariant spectrum of initial fluctuation as
\begin{eqnarray}P_\Psi(k)=A(k/H_0)^{n_s-1}=A\sim10^{-10},\quad n_s=1.
\end{eqnarray}

The wavelength of perturbation is $\lambda=2\pi a(\eta)/k$, where $k$ is the comoving wave number. During radiation-dominated epoch, perturbation keeps adiabatic on super horizon. While $\lambda$ entering horizon, perturbation begins to oscillate and decay because of the interaction with high temperature photons. During this epoch
\begin{eqnarray}\lambda=2\pi a(\eta)/k\sim ct\propto a^2 .\label{eq23}\end{eqnarray}
So perturbation with wave number $k$ enters horizon at time $t_k\propto a^2\propto k^{-2}$. In short, perturbation with smaller scale decays earlier. So the transfer function at very small scale satisfies $T(k)\propto t_k\propto k^{-2}$. While during matter domination, perturbation keeps increasing, so $T(k)$ tends to be a constant at large scale \cite{insert1}. We finally approximate the transfer function as
\begin{eqnarray}T^2(k)\simeq\frac{1}{1+\beta(k/k_{eq})^4}=\frac{1}{1+\beta(k\eta_{eq})^4} ,\end{eqnarray}
where $k_{eq}$ and $\eta_{eq}$ represent the wave number and conformal time at matter and radiation equality respectively. In Dodelson's book \cite{29} or Peacock's book \cite{30}, the authors list the analytical expression of the transfer function in CDM universe and we find that $T^2(k)$ with $\beta\simeq 2.7\times10^{-4}$ is consistent with analytical one grossly. In dealing with $C_\ell^{(3)}$, we consider some correction on the transfer function to fit with the analytical expression better at small scale.

To get the analytical result of the Hubble parameter power spectrum, we also need some appropriate approximations to calculate the integral
$\int{dxT^2(x)\frac{j_\ell(x)j_\ell(\xi x)}{x^\lambda}}$, where $j_\ell(x)$ is $\ell$th spherical Bessel function. We discuss this integral in two different cases.

(i) Case $\lambda >-1, \xi \leqslant 1 $. $\int{dx x^{-\lambda}{j_\ell(x)j_\ell(\xi x)}}$ is integrable according to Eq.~(\ref{eq:B4}). As $x$ increasing, both ${j_\ell(x)j_\ell(\xi x)}/{x^\lambda}$ and $T^2(x)$ decay very fast during $(0,\infty)$. Furthermore, $j_\ell(x)$ is dominated by the first peak which distributes near $x\simeq\ell$ and $j_\ell(x)j_\ell(\xi x)$ is almost always peaked at $\ell_\xi\equiv\frac{1}{2}\ell(1+1/\xi)$. Thus we assume
\begin{eqnarray}
    \int{dxT^2(x)\frac{j_\ell(x)j_\ell(\xi x)}{x^\lambda}}\simeq T^2(\ell_\xi)\int{dx\frac{j_\ell(x)j_\ell(\xi x)}{x^\lambda}}.\label{eq25} \nonumber \\ \end{eqnarray}
Comparing with numerical result, we find the condition $\xi<1$ is acceptable, but the result for $\xi=1$ is excellent with an error less than $0.5\%$ for $\ell<\ell_{pk}$, where $\ell_{pk}$ is the peak of the power spectrum. We mainly care about the Hubble parameter power spectrum i.e. $\xi=1$ rather than the correlation function i.e. $\xi<1$. Besides, it becomes worse while $\ell$ increases for $\xi=1$, and the value is about $5\%$ overvalued at $\ell_{max}\simeq 3\ell_{pk}$.

(ii) Case $\lambda =-1, \xi = 1$. The integral $\int{dx xj^2_\ell(x)}=\infty$, so approximation in (i) could not be used any more. As mentioned in (i), because $j_\ell(x)$ is dominated by the first peak, we assume $\int{dxxT^2(x)j^2_\ell(x)}\simeq \ell I^2_\ell T^2(\ell)$, where $I^2_\ell$ is a value dependent on $\ell$. Numerical computation shows $I^2_\ell\simeq 1.6/\ell^2$, so
\begin{eqnarray}\int{dxxT^2(x)j^2_\ell(x)}\simeq \ell I^2_\ell T^2(\ell). \label{eq26}  \end{eqnarray}
In fact, this method is just the Limber approximation by replacing 1.6 with $\pi/2$ \cite{luminosity}. However this approximation is not reasonable for $\ell=1,2$, so we plot spectrums beginning from $\ell=3$. In addition, we find this method is effective with small error almost for all $\ell$.

The integral over k from $0$ to $\infty$ in method (i) induces the terms of hypergeometric functions which have been listed in Appendix C and method (ii) obtains a concise analytical expression. Our approximate result is a little underestimated compared to the numerical result, but no more than $5\%$ during $\ell_{pk}<\ell<\ell_{max}$, and they are consistent quite well during $3<\ell<\ell_{max}$. More details are given in Appendix A. In this paper we only focus on the analytical results contributed by different $C_\ell^{(i)}$ but don't have so much interest in the numerical results which in turn can be used to test the accuracy of the analytical ones\cite{note}.

\begin{figure}
  \centering
  \includegraphics[width=3.0in,height=2.7in]{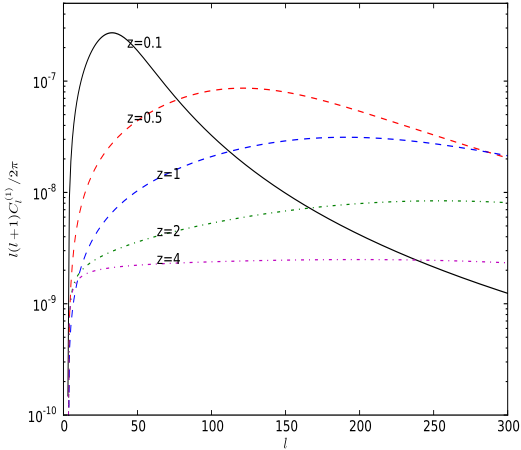}
  \caption{Power spectrums of $\ell(\ell+1)C_\ell^{(i)}(z,z)/2\pi$ for z=0.1, 0.5 1, 2 and 4 from top to bottom.}\label{C1}
\end{figure}

\begin{figure}
  \centering
  \includegraphics[width=3.0in,height=2.7in]{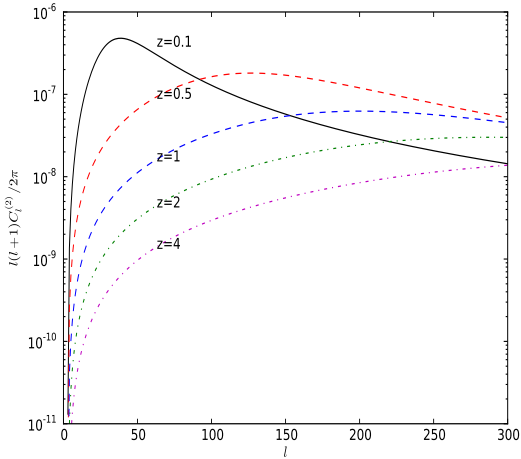}
  \caption{Same as Fig.~\ref{C1}, but for the contribution of $\ell(\ell+1)C_\ell^{(i)}(z,z)/2\pi$. The line style is same with Fig.~\ref{C1}. }\label{C2}
\end{figure}

\begin{figure}
  \centering
  \includegraphics[width=3.0in,height=2.7in]{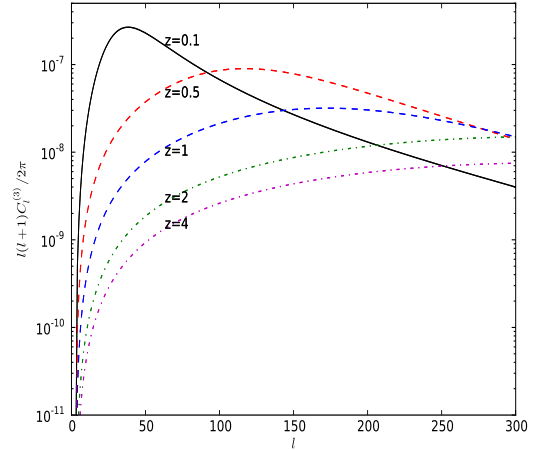}
  \caption{Same as Fig.~\ref{C1}, but for the contribution of the Doppler term $\ell(\ell+1)C_\ell^{(i)}(z,z)/2\pi$ for z=0.1, 0.5 1, 2 and 4 respectively.}\label{C3}
\end{figure}

In Figs.~\ref{C1}-\ref{C3}, we show the power spectrum $\ell(\ell+1)C_\ell^{(i)}/2\pi$ at different redshift $z$. Farther calculation shows the peak of the spectrums corresponds to scale $k_{eq}$ which enters horizon at conformal time $\eta_{eq}$. According to the relation ${\pi}/{\ell}={\lambda}/{\chi}$, a small scale at low redshift builds up to a very large scale at high redshift for the same angular deviation, where $\lambda$ can be found in Eq.~(\ref{eq23}) and $\chi$ is the angular diameter distance. So the power spectrum at low redshift turns over at small $\ell$. Another interesting phenomenon is that $C_\ell^{(2)}(z,z)\simeq 2C_\ell^{(3)}(z,z)\simeq2C_\ell^{(1)}(z,z)$. We find that $C_\ell^{(\eta\eta')}(z,z)$, $C_\ell^{(\eta n')}(z,z)$, $C_\ell^{(n\eta')}(z,z)$ and $C_\ell^{(nn')}(z,z)$ have the same expression when applying the conformal position $\textbf{x}=\textbf{x}_0+\textbf{n}(\eta_0-\eta)$ to derive the Hubble parameter power spectrum. To be convenient, we name the four terms above as Doppler term $C_\ell^{(Dop)}$ together which dominate in power spectrums. Mathematically speaking, the reason why $C_\ell^{(Dop)}$ dominates is that $\int{dx\frac{j_\ell(x)j_\ell(x)}{x}}\sim\int{dx j'_\ell(x)j_\ell(x)}\ll\int{dx xj'_\ell(x)j'_\ell(x)}$. Physically speaking, peculiar velocity, caused by gravitational potential, contributes to the fluctuation primarily.

\begin{figure}
  \centering
  \includegraphics[width=3.0in,height=2.5in]{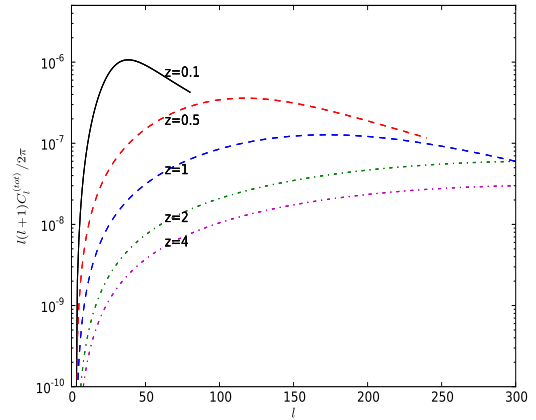}
  \caption{The total Hubble parameter power spectrum $C_\ell^{(tot)}(z,z)$. Since spectrums at $\ell>\ell_{max}$ are not reliable, we only plot spectrums for z=0.2 and 0.5 on $\ell<\ell_{max}$.}\label{Ctot}
\end{figure}

\begin{figure}
  \centering
  \includegraphics[width=3.0in,height=2.5in]{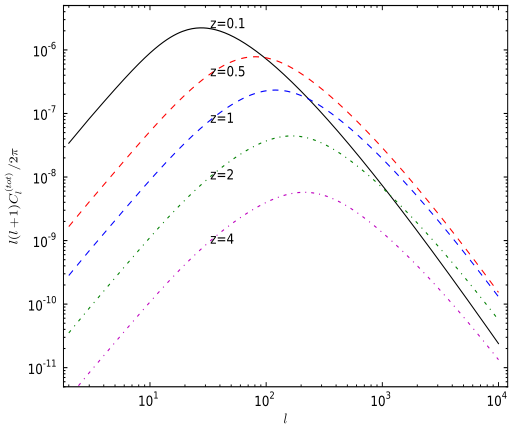}
  \caption{The total Hubble parameter power spectrum $C_\ell^{(tot)}(z,z)$ with log correlation.}\label{Ctot_log}
\end{figure}

\begin{figure}
  \centering
  \includegraphics[width=3.0in,height=2.5in]{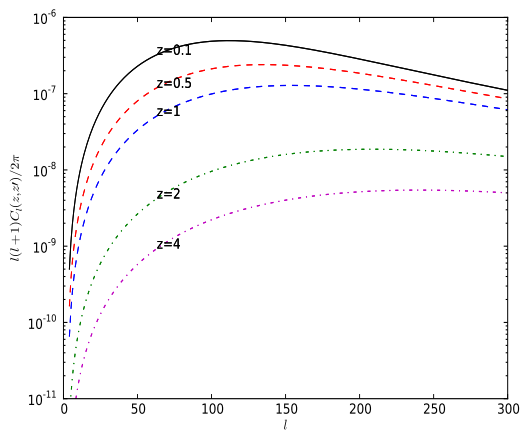}\\
  \caption{Contribution of correlation function $\ell(\ell+1)C_\ell(z,z')/2\pi$ is plotted for $z'=1$ and z=0.1, 0.5 1, 2, 4 from top to bottom.}\label{corre_1}
\end{figure}

\begin{figure}
  \centering
  \includegraphics[width=3.0in,height=2.5in]{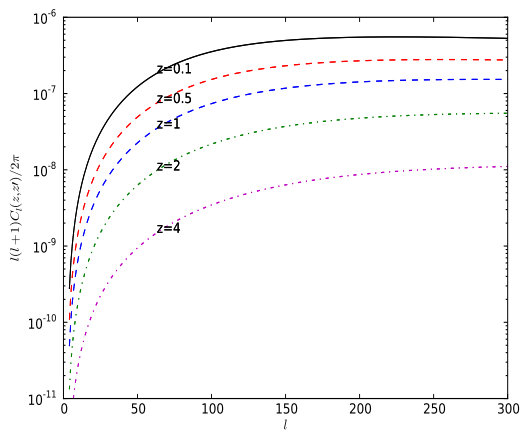}\\
  \caption{Same as Fig.~\ref{corre_1}, but  for $z'=2$.}\label{corre_2}
\end{figure}

Fig.~\ref{Ctot} shows the sum of $C_\ell^{(i)}$
$$\ell(\ell+1)\Big[\sum_i{C_\ell^{(i)}(z,z)} \Big]/2\pi.$$
The spectrum at $z=0.1$ is plotted during the internal $3\leqslant\ell\leqslant300$. However, as we discussed above, result during $3\leqslant\ell\leqslant\ell_{max}\simeq80$ is reliable, while $\ell_{max}\simeq250$ for $z=0.5$. Other spectrums for $z\geqslant1$ have larger $\ell_{max}$ and they are all larger than $300$. Spectrum at low redshift has a shaper distribution, which comes from the growth of the gravitational potential $\Psi(\eta,\textbf{x})$. However, if we want to analysis the properties on very small scale, such calculation is not reasonable any more. Fortunately, Appendix A provides an effective method to calculate the spectrum at higher $\ell$, which is used to plot Fig.~\ref{Ctot_log}. So Fig.~\ref{Ctot} shows the Hubble parameter power spectrum on large scale without the log correlation while Fig.~\ref{Ctot_log} shows the total spectrums on both very large and very small scale. And $\ell(\ell+1)C^{(3)}_\ell$ with the log correlation decays like $\ell^{-3}\textrm{log}^2\ell$. Another interesting feature is the behavior of correlation function $C_\ell^{(tot)}(z,z')$ for $z\neq z'$ as a function of $\ell$. Fig.~\ref{corre_1} shows $C_\ell(z,z')$ for $z=1$ and Fig.~\ref{corre_2} shows $C_\ell(z,z')$ for $z=2$. We can see that $C_\ell(z,z')$ tends to be independent of $\ell$ for large $z$ or $z'$(like $z\geqslant 2$).

 \section{\label{sec5}CONCLUSIONS AND DISCUSSION}
In this paper we derive the power spectrum and correlation function of the Hubble parameter. The fluctuation of the Hubble parameter comes from the perturbation of background metric (or gravitational potential). We obtain the semi-analytical expression with some approximation on the Bardeen transfer function. We found $C_\ell^{(Dop)}$ is dominant in $C_\ell^{(tot)}$. Namely, peculiar velocity contributes most to the fluctuation.

Using the data of $C_\ell^{(3)}$, we roughly get the total fluctuation of the Hubble parameter at any redshift between $z=0.1$ and $z=4$:
\begin{eqnarray*}\langle\delta_H(\textbf{n},z)\delta_H(\textbf{n}',z')\rangle=\sum_\ell \frac{2\ell+1}{4\pi}C_\ell\sim10^{-6}\ll 1.\end{eqnarray*}
It is almost 4 orders of magnitude larger than the CMB power spectrum. This result indicates that second order perturbation seems not responsible for the accelerated expansion of our universe \cite{13,14,15}. It's worth to mention that our results are obtained within the frame of the linear perturbation theory and have not considered the correction of the transfer function due to nonlinearity. Additionally, we calculate the power spectrum in CDM universe. In $\Lambda \textrm{CDM}$, both the transfer function and the growth function on any scale are smaller compared to CDM universe. So Hubble parameter has a larger fluctuation in CDM universe, which in turn becomes a potential observational tool to study the dark energy.

\begin{figure}
  \centering
  \includegraphics[width=3.0in,height=2.5in]{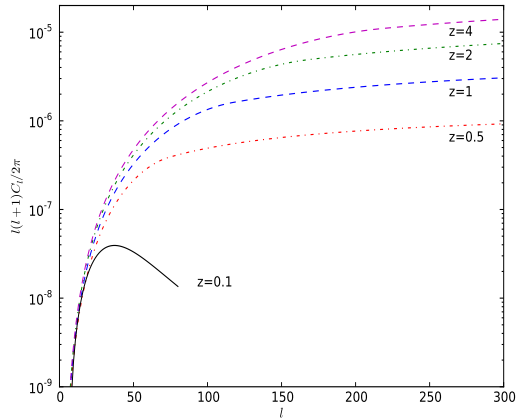}
  \caption{The luminosity distance power spectrums calculated by Camille Bonvin et al \cite{26}. Total $\ell(\ell+1)C_\ell(z,z)/2\pi$ for z=0.1, 0.5 1, 2 and 4 is shown from top to bottom. Combining this spectrum with our result, we reckon that this is a new way to constrain cosmological parameters and to study the dark energy.}\label{luminosity}
\end{figure}

We roughly restore the luminosity distance power spectrums in Fig.~\ref{luminosity} given by Camille Bonvin et al. \cite{26}. We can see the luminosity distance at high redshift has larger fluctuation since photons from distant supernova or galaxy travel through more gravitational potential. But our Hubble parameter power spectrums at low redshift has larger fluctuation because of the growth of gravitational potential. So combination of both tools together may be a potential method to determine cosmological parameters, to constrain the global anisotropy and even to study the inhomogeneity of dark energy.

Finally, we discuss the measurement of this perturbed quantity. First of all, such fluctuation could be detected. The Hubble parameter as a direct way to measure the cosmic expansion is related to cosmological distance, especially the luminosity distance \cite{luminosity}. We get the perturbed Hubble parameter by derivative of perturbed luminosity distance (see Appendix B). As the luminosity distance's fluctuation is widely studied \cite{24,25,26,insert2,luminosity2}, we believe that the Hubble parameter's fluctuation can be detected in this way, which invokes us to do the deeper research. On the other hand, Eq.~(\ref{velocity}) provides a method to detect such fluctuation from velocity field. As shown in Fig.~\ref{Ctot} and Fig.~\ref{luminosity}, the Hubble parameter has a larger fluctuation than the luminosity distance at $z\lesssim 0.2$. We can get enough survey on velocity of galaxies, i.e. $\textbf{v}$ in Eq.~(\ref{insert5}), at the same redshift $z<0.2$ \cite{insert4}. Then recession velocity is obtained by deducting the peculiar velocity $\textbf{v}_p$ that dominates in the luminosity distance power spectrum from $\textbf{v}$. If recession velocity is not a constant, it means that the fluctuation of the Hubble parameter may exist. In a word, inhomogeneity of constant-$\eta$ hypersurface gives rise to fluctuation of local expansion rate. That's to say, many distant luminous celestial bodies on geodesic motion in perturbed metric may contain such physical quantity, such as supernovas \cite{supernova}. Second, spectrums tend to peak at small $\ell$ as $z$ decreases. When $z\ll 1$, spectrum distributes shapely at extremely small $\ell$(like $\ell$=1). Such dipole is just the motion of ourselves which can be easily detected by measuring the CMB dipole. This can be a direct evidence for the existence of the Hubble parameter's fluctuation. Then, the gravitational wave may also be a potential method to detect such quantity \cite{insertx}.

\begin{acknowledgments}
We thank Xi Yang and Jian-Chuan Zheng for useful and stimulating discussions. This work was supported by the National Science Foundation of China (Grants No. 11173006), the Ministry of Science and Technology National Basic Science program (project 973) under grant No. 2012CB821804.
\end{acknowledgments}

\appendix
\begin{widetext}
\section{INTEGRALS AND FORMULAS}
To derive the expression of $C_{\ell}^{(i)}(z,z')$, we take full advantage of approximation (i) and (ii) in Sec. \ref{sec4}. The fact is always used that the Bardeen transfer function is independent on $\eta$. Furthermore, the initial scale-invariant spectrum is $P_\Psi(k)=A\simeq10^{-10}$. We also need to define some other parameters, $b_z\equiv\frac{\eta}{\eta_0}=\frac{1}{\sqrt{1+z}}$, $x=k(\eta_0-\eta)$. And note that $x'=\frac{1-b_{z'}}{1-b_z}x$. Then the transfer function becomes
$$T^2(x)=\frac{1}{1+\alpha_z x^4}, \eqno(A1)$$
where $\alpha_z=\beta(\frac{b_{eq}}{1-b_z})^4$, and $b_{eq}=\frac{1}{\sqrt{1+z_{eq}}}\simeq0.01$. This expression will be used immediately in the calculation of $C_{\ell}^{(1)}$ and $C_{\ell}^{(2)}$.

\subsection{$\textbf{C}_\ell^\textbf{(1)}$}
\begin{eqnarray}
C_{\ell}^{(1)}(z,z')&=&\frac{2}{\pi}\Big\{\frac{1}{4}\int{\frac{d k}{k}T^2(k)P_\Psi(k)j_{\ell}(k(\eta_0-\eta))j_{\ell}(k(\eta_0-\eta '))}\nonumber \\
    & &-\frac{1}{2\mathcal{H}(\eta)}\frac{\partial}{\partial \eta}\int{\frac{d k}{k}T^2(k)P_\Psi(k)j_{\ell}(k(\eta_0-\eta))j_{\ell}(k(\eta_0-\eta '))}\nonumber\\
    & &-\frac{1}{2\mathcal{H}(\eta ')}\frac{\partial}{\partial \eta '}\int{\frac{d k}{k}T^2P_\Psi(k)(k)j_{\ell}(k(\eta_0-\eta))j_{\ell}(k(\eta_0-\eta '))}\nonumber\\
    & &+\frac{1}{\mathcal{H}(\eta)}\frac{1}{\mathcal{H}(\eta ')}\frac{\partial^2}{\partial \eta \partial \eta '}\int{\frac{d k}{k}T^2(k)P_\Psi(k)j_{\ell}(k(\eta_0-\eta))j_{\ell}(k(\eta_0-\eta '))}\Big\}\nonumber\\
    &=&A\Big\{\frac{1}{4}\int{\frac{d x}{x^2}T^2(x)J_{\ell+1/2}(x)J_{\ell+1/2}(\xi x))}-\frac{b_z}{2}\frac{\partial}{\partial b_z}\int{\frac{d x}{x^2}T^2(x)J_{\ell+1/2}(x)J_{\ell+1/2}(\xi x))}\nonumber\\
    & &-\frac{b_{z'}}{2}\frac{\partial}{\partial b_{z'}}\int{\frac{d x}{x^2}T^2(x)J_{\ell+\frac{1}{2}}(x)J_{\ell+1/2}(\xi x))}+b_z b_{z'}\frac{\partial ^2}{\partial b_z \partial b_{z'}}\int{\frac{d x}{x^2}T^2(x)J_{\ell+\frac{1}{2}}(x)J_{\ell+1/2}(\xi x))} \Big\}\nonumber\\
    &=&\frac{A\Gamma(\ell)}{2^4\Gamma(\ell+\frac{3}{2})\Gamma(\frac{3}{2})}\frac{1}{1+\alpha_z \ell_\xi^4}\frac{(1-b_{z})^{\ell}(1-b_{z '})^{\ell}}{(2-b_{z}-b_{z'})^{2\ell}}F\Big(\ell,\ell+1;2\ell+2;\frac{4(1-b_{z})(1-b_{z'})}{(2-b_{z}-b_{z'})^2}\Big)\nonumber \\
    & &+\frac{1}{1+\alpha_z \ell_\xi^4}\frac{A b_z\Gamma(\ell)}{2^4\Gamma(\ell+\frac{3}{2})\Gamma(\frac{3}{2})}\times\Big[\frac{\ell(1-b_{z})^{\ell-1}(1-b_{z '})^{\ell}(b_z-b_{z'})}{(2-b_{z}-b_{z'})^{2\ell+1}}F\Big(\ell,\ell+1;2\ell+2;\frac{4(1-b_{z})(1-b_{z'})}{(2-b_{z}-b_{z'})^2}\Big)\nonumber \\
   & &+\frac{2\ell(1-b_{z})^{\ell}(1-b_{z '})^{\ell+1}(b_{z'}-b_z)}{(2-b_{z}-b_{z'})^{2\ell+3}}
    F\Big(\ell+1,\ell+2;2\ell+3;\frac{4(1-b_{z})(1-b_{z'})}{(2-b_{z}-b_{z'})^2}\Big)\Big]+b_z\Longleftrightarrow b_{z'}\nonumber \\
    & &+\frac{Ab_z b_{z'}}{1+\alpha_z \ell_\xi^4}\times\Big[\frac{ \Gamma(\ell+1)}{2^3\Gamma(\ell+\frac{3}{2})\Gamma(\frac{3}{2})}\frac{(1-b_{z})^{\ell-1}(1-b_{z '})^{\ell-1}}{(2-b_{z}-b_{z'})^{2\ell+2}}\times\nonumber \\
    & &[(2\ell+1)(1-b_{z'})(b_z-b_{z'})-(2-b_z-b_{z'})(\ell(b_z-b_{z'})-b_{z'}+1)]
    F\Big(\ell,\ell+1;2\ell+2;\frac{4(1-b_{z})(1-b_{z'})}{(2-b_{z}-b_{z'})^2}\Big)\nonumber \\
    & &-\frac{ \Gamma(\ell+1)}{2^2\Gamma(\ell+\frac{3}{2})\Gamma(\frac{3}{2})}\frac{(1-b_{z})^{\ell}(1-b_{z '})^{\ell}(b_{z}-b_{z'})}{(2-b_{z}-b_{z'})^{2\ell+4}}
     F\Big(\ell+1,\ell+2;2\ell+3;\frac{4(1-b_{z})(1-b_{z'})}{(2-b_{z}-b_{z'})^2}\Big)\nonumber \\
    & &-\frac{\Gamma(\ell+3)}{2\Gamma(\ell+\frac{5}{2})\Gamma(\frac{3}{2})}\frac{(1-b_{z})^{\ell+1}(1-b_{z '})^{\ell+1}(b_{z'}-b_z)^2}{(2-b_{z}-b_{z'})^{2\ell+6}}
     F\Big(\ell+2,\ell+3;2\ell+4;\frac{4(1-b_{z})(1-b_{z'})}{(2-b_{z}-b_{z'})^2}\Big) \Big]\label{eq:A1}.
\end{eqnarray}
where Eqs.~(\ref{eq25}) and ~(\ref{eq:B3}) -~(\ref{eq:B7}) have been used. And we assume that $z>z'$ i.e. $\xi<1$ (if instead, we just need to reverse $z$ and $z'$).

\subsection{$\textbf{C}_\ell^\textbf{(2)}$}
$C_\ell^{(2)}$ is consisted of four terms:
$$C_\ell^{(2)}(z,z')=C_\ell^{(n\Psi ')}(z,z')+C_\ell^{(\Psi n')}(z,z')+C_\ell^{(\eta\Psi ')}(z,z')+C_\ell^{(\Psi \eta')}(z,z'),$$
According to Eq.~(\ref{spectrum4}), the expression of $C_\ell^{(n\Psi')}$ is
\begin{eqnarray}
    C_\ell^{(n\Psi ')}(z,z')&=&\frac{1}{2}\frac{2}{\pi}\frac{1}{\mathcal{H}(\eta)}\int{dk T^2(k)P_{\Psi}(k)j_\ell(k(\eta_0-\eta))j'_\ell(k(\eta_0-\eta '))}\nonumber \\
    &=&\frac{A}{4}\frac{b_z}{1-b_z}\int{dx \frac{1}{1+\alpha_z x^4} \Big[ \frac{1}{x}J_{\ell-1/2}(x)J_{\ell+1/2}(\xi x)-\frac{\ell+1}{x^2}J_{\ell+1/2}(x)J_{\ell+1/2}(\xi x) \Big]}\nonumber \\
    &=&\frac{A}{4}\frac{b_z}{1-b_z}\frac{\Gamma(\ell)\xi^\ell}{2^2\Gamma(\ell+\frac{3}{2})\Gamma(\frac{3}{2})}\frac{1}{1+\alpha_z \ell_{\xi}^4}\times\nonumber \\
     & &\Big[(\ell+1)F(\ell,-\frac{1}{2};\ell+\frac{3}{2};\xi^2)-F(\ell,\frac{1}{2};\ell+\frac{3}{2};\xi^2) \Big]. \label{eq:A2}
\end{eqnarray}
Similarly, we can get the expression of $C_\ell^{(\Psi n')}(z,z')$
\begin{eqnarray}
    C_\ell^{(\eta\Psi ')}(z,z')&=&-\frac{A}{2\pi}\frac{b_z}{1-b_z}b_{z'}\frac{\partial}{\partial b_{z'}}\int{dx T^2(x)j'_\ell(x)j_\ell(\xi x)}\nonumber \\
    &=&\frac{A}{4}\frac{b_z b_{z'}\xi^{-1/2}}{(1-b_z)^2}\Big\{\int{dx\frac{1}{1+\alpha_z x^4}\frac{(\ell+1)^2}{\xi x^2}J_{\ell+1/2}(x)J_{\ell+1/2}(\xi x)}-\nonumber \\
     & &\int{dx\frac{1}{1+\alpha_z x^4}\frac{\ell+1}{x}J_{\ell+1/2}(x)J_{\ell-1/2}(\xi x)}-\int{dx\frac{1}{1+\alpha_z x^4}\frac{\ell+1}{\xi x}J_{\ell-1/2}(x)J_{\ell+1/2}(\xi x)}\nonumber \\
     & &+\int{dx\frac{1}{1+\alpha_z x^4}J_{\ell-1/2}(x)J_{\ell-1/2}(\xi x)} \Big\}\nonumber \\
     &=&\frac{A}{4}\frac{b_z b_{z'}}{(1-b_z)^2}\frac{1}{1+\alpha_z \ell_{\xi}^4}\times\Big[\frac{\xi^{\ell-1}(\ell+1)^2\Gamma(\ell)}{2^2\Gamma(\ell+\frac{3}{2})\Gamma(\frac{3}{2})}F(\ell,-\frac{1}{2};\ell+\frac{3}{2};\xi^2)-
     \nonumber\\
     & &\frac{\xi^{\ell-1}(\ell+1)\Gamma(\ell)}{2\Gamma(\ell+\frac{1}{2})\Gamma(\frac{3}{2})}F(\ell,-\frac{1}{2};\ell+\frac{1}{2};\xi^2)-
      \frac{\xi^{\ell-1}(\ell+1)\Gamma(\ell)}{2\Gamma(\ell+\frac{3}{2})\Gamma(\frac{1}{2})}F(\ell,\frac{1}{2};\ell+\frac{3}{2};\xi^2)+\nonumber\\
     & &\frac{\xi^{\ell-1}\Gamma(\ell)}{\Gamma(\ell+\frac{1}{2})\Gamma(\frac{1}{2})}F(\ell,\frac{1}{2};\ell+\frac{1}{2};\xi^2)  \Big] \qquad\qquad (\xi<1).\label{eq:A3}
\end{eqnarray}
We can also get $C_\ell^{(\Psi \eta')}(z,z')$ in the same way. However, the computation above is only for $\xi<1$. When $\xi=1$ the integral $\int{dx J^2_\ell(x)}=\infty$, so we have to deal with it by Eq.~(\ref{eq26}). Then,
$$\frac{A}{4}\frac{b_z b_{z'}}{(1-b_z)^2}\int{dx\frac{x}{1+\alpha_z x^4}j_{\ell-1}(x)j_{\ell-1}(\xi x)}\simeq \frac{A}{2\pi}\frac{b_z b_{z'}}{(1-b_z)^2}\frac{\ell-1}{1+\alpha_z (\ell-1)^4}I^2_{\ell-1} \qquad (\xi=1).$$
 Finally, the complete expression of $C_\ell^{(2)}(z,z')$ is
\begin{eqnarray}
C_\ell^{(2)}(z,z')&=&\frac{Ab_z\Gamma(\ell)(1-b_{z'})^\ell}{2^4\Gamma(\ell+\frac{3}{2})\Gamma(\frac{3}{2})
    (1-b_z)^{\ell+1}}\frac{1}{1+\alpha_z\ell_{\xi}^4}\Big[F(\ell,\frac{1}{2};\ell+\frac{3}{2};\xi^2)
    -(\ell+1)F(\ell,-\frac{1}{2};\ell+\frac{3}{2};\xi^2) \Big]\nonumber\\
& &+\frac{Ab_{z'}\Gamma(\ell)(1-b_{z'})^{\ell-1}}{2^4\Gamma(\ell+\frac{3}{2})\Gamma(\frac{3}{2})(1-b_z)^{\ell}}
    \frac{1}{1+\alpha_{z'}\ell_{\xi'}^4}\Big[F(\ell,-\frac{1}{2};\ell+\frac{1}{2};\xi^2)-(\ell+1)
    F(\ell,-\frac{1}{2};\ell+\frac{3}{2};\xi^2) \Big]\nonumber\\
& &+\frac{A b_z b_{z'}}{4}\frac{(1-b_{z'})^{\ell-1}}{(1-b_z)^{\ell+1}}\Big(\frac{1}{1+\alpha_z\ell_{\xi}^4}
    +\frac{1}{1+\alpha_{z'} \ell_{\xi'}^4} \Big)\times\nonumber\\
& &\Big[\frac{(\ell+1)^2\Gamma(\ell)}{2^2\Gamma(\ell+\frac{3}{2})\Gamma(\frac{3}{2})}
    F(\ell,-\frac{1}{2};\ell+\frac{3}{2};\xi^2)
    -\frac{(\ell+1)\Gamma(\ell)}{2\Gamma(\ell+\frac{1}{2})\Gamma(\frac{3}{2})}
    F(\ell,-\frac{1}{2};\ell+\frac{1}{2};\xi^2)\nonumber\\
& &-\frac{(\ell+1)\Gamma(\ell)}{2\Gamma(\ell+\frac{3}{2})\Gamma(\frac{1}{2})}
    F(\ell,\frac{1}{2};\ell+\frac{3}{2};\xi^2)+\frac{\Gamma(\ell)}{\Gamma(\ell+\frac{1}{2})\Gamma(\frac{1}{2})}
    F(\ell,\frac{1}{2};\ell+\frac{1}{2};\xi^2)  \Big] \qquad (\xi<1),\label{eq:A4}
\end{eqnarray}
where $\alpha_{z'}=\beta(\frac{b_{eq}}{1-b_{z'}})^4$, $\ell_{\xi'}\equiv\ell(1+\xi)/2$ and
\begin{eqnarray}
 C_\ell^{(2)}(z,z)&=&-\frac{Ab_z\Gamma(\ell)(1-b_{z'})^\ell}{2^3\Gamma(\ell+\frac{3}{2})\Gamma(\frac{3}{2})(1-b_z)^{\ell+1}}\frac{1}{1+\alpha_z \ell^4}\Big[2(\ell+1)F(\ell,-\frac{1}{2};\ell+\frac{3}{2};\xi^2)\nonumber\\
 & &-F(\ell,\frac{1}{2};\ell+\frac{3}{2};\xi^2)-F(\ell,-\frac{1}{2};\ell+\frac{1}{2};\xi^2) \Big]
 +\frac{A b_z^{2}}{2(1-b_z)^2}\frac{1}{1+\alpha_z \ell^4}\times\nonumber\\
 & &\Big[\frac{(\ell+1)^2\Gamma(\ell)}{2^2\Gamma(\ell+\frac{3}{2})\Gamma(\frac{3}{2})}
    F(\ell,-\frac{1}{2};\ell+\frac{3}{2};\xi^2)-
    frac{(\ell+1)\Gamma(\ell)}{2\Gamma(\ell+\frac{1}{2})\Gamma(\frac{3}{2})}
    F(\ell,-\frac{1}{2};\ell+\frac{1}{2};\xi^2)\nonumber\\
& &-\frac{(\ell+1)\Gamma(\ell)}{2\Gamma(\ell+\frac{3}{2})\Gamma(\frac{1}{2})}
    F(\ell,\frac{1}{2};\ell+\frac{3}{2};\xi^2)  \Big]
    +\frac{2A}{\pi}\frac{b_z^{2}}{(1-b_z)^2}\frac{\ell-1/2}{1+\alpha_z (\ell-1)^4}I^2_{\ell-1} \qquad(\xi=1).\label{eq:A5}
\end{eqnarray}
In the formulas above, Eqs.~(\ref{eq25}), ~(\ref{eq26}) and ~(\ref{eq:B2}) -~(\ref{eq:B5}) have been employed.
\end{widetext}
 \subsection{$\textbf{C}_\ell^\textbf{(3)}$}
 The transfer function in Eq.~(\ref{eq:A1}) is not consistent very well at small scale. To solve this problem, we need to do some correction on the transfer function. One method is using the numerical result \cite{30}, however, this way is something overcorrect. Luckily, Camille Bonvin et al. \cite{26} have provided an applied method and they call it log correction
\begin{eqnarray} T^2(x)=\frac{1}{1+\alpha_z x^4/\textrm{log}^2(1+\frac{7.8\times10^{-4}}{1-b_z}x)}. \end{eqnarray}
According to Eq.~(\ref{eqC3}), the expression of $C_\ell^{(3)}$ is
\begin{eqnarray}C_\ell^{(3)}(z,z')=\frac{2A\xi b_z b_{z'}}{\pi (1-b_{z})(1-b_{z'})}\int{dx xT^2(x)j'_\ell(x)j'_\ell(\xi x)}. \nonumber\\ \end{eqnarray}
Using Eq.(B2) and approximation (ii), and focusing only on the power spectrum($\xi=1$), we have
\begin{eqnarray}C_\ell^{(3)}(z,z)=&&\frac{2b_z^{2}}{\pi(1-b_{z})^2}\times\Big\{(\ell-1)T^2(\ell-1)I^2_{\ell-1}\nonumber\\
&&-\frac{1}{2}(\ell+1)T^2(\ell-1)I^2_{\ell-1}-\frac{1}{2}(\ell+1)T^2(\ell)I^2_{\ell}\nonumber\\&&+(\ell+1)^2T^2(\ell)I^2_{\ell}/\ell  \Big\}.
\end{eqnarray}
The factor $1/2$ comes from $j_{\ell-1}(x)j_{\ell}(x)$ term in which the peak is lower and negative parts arise. However, $C_\ell^{(3)}$ is a little overvalued by this approach at high $\ell$. So we must make some correction on the expression above. Comparing with the numerical results, we find such expression is consistent with it reasonably
\begin{eqnarray}C_\ell^{(3)}&&(z,z)=\frac{2b_z^{2}}{\pi (1-b_{z})^2}\frac{1}{0.5(1-b_z)^2+0.6b_z^2}\frac{2.5}{1+100/\ell^{1.5}}\nonumber\\
\times&&\Big\{(\ell-1)T^2(\ell-1)I^2_{\ell-1}-\frac{1}{2}(\ell+1)T^2(\ell-1)I^2_{\ell-1}\nonumber\\
&&-\frac{1}{2}(\ell+1)T^2(\ell)I^2_{\ell}+(\ell+1)^2T^2(\ell)I^2_{\ell}/\ell  \Big\}\label{eq:A9}.
\end{eqnarray}

\begin{figure}
  \centering
  \includegraphics[width=3.0in,height=3.0in]{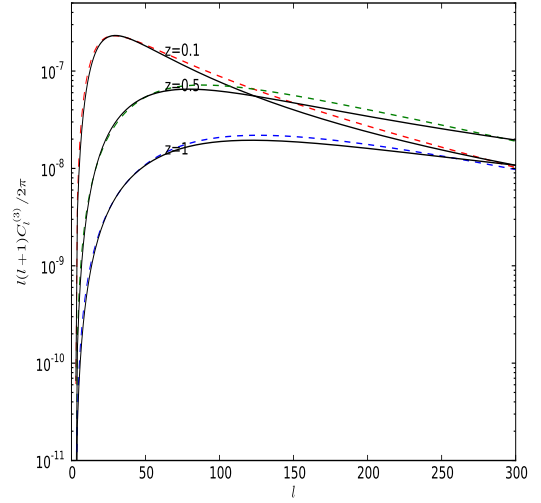}
  \caption{The approximate results (Black solid lines) of $\ell(\ell+1)C_\ell^{(3)}(z,z)/2\pi$ given by Eq.(A9) are compared to the numerical results (three dash lines) for z=0.1, 0.5 and 1 respectively. The numerical results have also been used to validate method (i) and (ii) in Sec. \ref{sec4}.}\label{Cm3}
\end{figure}

In Fig.~\ref{Cm3}, we plot the approximate power spectrums and the numerical results for $z=0.1$, $z=0.5$ and $z=1$. For comparison, this result is about $5\%$ smaller than the numerical treatment. Fig.~\ref{C3} uses the the analytical results in Eq.~(\ref{eq:A5}) without correction while Fig.~\ref{Ctot_log} uses the fitting results in Eq.~(\ref{eq:A9}) with the correction. We can also conclude that $C_\ell^{(3)}$ decays like $\ell^{-3}\textrm{log}^2\ell$ when $\ell\gg\ell_{max}$. Again, such method provides an effective method to calculate the spectrum at very large $\ell$ i.e. on very small scale.

\section{PERTURBED HUBBLE PARAMETER VIA LUMINOSITY DISTANCE}
The Hubble parameter is normally measured by the derivative of the quantity of the luminosity distance
\begin{eqnarray}H(z)=\frac{1}{\frac{d}{dz}[d_L(z)/(1+z)]}. \label{eq:C1} \end{eqnarray}
In a perturbed universe, the Hubble parameter at redshift $\tilde{z}$ along direction $\textbf{n}$ is
\begin{eqnarray}H(\tilde{z},\textbf{n})=\frac{1}{\frac{d}{d\tilde{z}}[(1+\tilde{z})^{-1}d_L(\tilde{z},\textbf{n})]}. \label{eq:C2} \end{eqnarray}
However $\tilde{z}$ is not measurable. The measurable quantity instead is $z=\tilde{z}+\delta z$, where
\begin{eqnarray}
    \delta z=& &(1+z)\Big[\Psi(\eta_0)-\Psi(\eta)+2\int^{\eta}_{\eta_0}{d\eta \dot{\Psi}}\nonumber\\
    & &+(\textbf{v}_0-\textbf{v})\cdot\textbf{n} \Big] \equiv (1+z)\delta \bar{z}. \label{eq:C3}
\end{eqnarray}
Meanwhile, the perturbed luminosity distance has the relation \cite{insert2}
\begin{eqnarray}
    d_L(\tilde{z},\textbf{n})=d_L(z,\textbf{n})-\frac{d}{d\tilde{z}}d_L(z,\textbf{n})\delta z \label{eq:C4}
\end{eqnarray}
and
\begin{eqnarray}
    \frac{d}{d\tilde{z}}d_L(z,\textbf{n})=(1+z)^{-1}d_L+\mathcal{H}^{-1}+\textrm{first order}, \label{eq:C5}
\end{eqnarray}
where $\mathcal{H}$ is the conformal Hubble parameter.

With the above, we have
\begin{eqnarray}
    H(\tilde{z},\textbf{n})&=&\Big\{\frac{dz}{d\tilde{z}}\frac{d}{dz}\big[(1+z-\delta z)^{-1}\nonumber\\
        & &\quad\quad\quad\quad\big(d_L(z,\textbf{n})-\frac{d}{d\tilde{z}}d_L(z,\textbf{n})\delta z\big)\big]\Big\}^{-1} \nonumber\\
    &=&\Big\{\frac{d}{dz}\big[(1+z)^{-1}\big(1-(1+z)^{-1}\delta z\big)\nonumber\\
        & &\quad\quad\quad\quad\big(d_L-\frac{d}{d\tilde{z}}d_L\delta z\big) \big] \Big\}^{-1}\Big\{\frac{dz}{d\tilde{z}}\Big\}^{-1} \nonumber\\
    &=&\Big\{\frac{d}{dz}\big[(1+z)^{-1}d_L\big]-\frac{d}{dz}\big[(1+z)^{-1}d_L\big]\delta\bar{z}\nonumber\\
        & &\quad\quad\quad\quad-\frac{d}{dz}\big[(1+z)^{-1}d_L+\nonumber\\
        & &\quad\quad\quad\quad\mathcal{H}^{-1}\big]\delta\bar{z} \Big\}^{-1} \Big\{\frac{dz}{d\tilde{z}}\Big\}^{-1} \nonumber\\
    &=&\Big\{\frac{d}{dz}\big[(1+z)^{-1}d_L\big]-2\frac{d}{dz}\big[(1+z)^{-1}d_L\big]\delta\bar{z}\nonumber\\
        & &\quad\quad\quad-\frac{d}{dz}\big[(1+z)H^{-1}(z)\big]\delta\bar{z}\Big\}^{-1}  \Big\{\frac{dz}{d\tilde{z}}\Big\}^{-1}\nonumber\\
    &=&\Big\{H^{-1}(z)-3H^{-1}(z)\delta\bar{z}\nonumber\\
        & &\quad\quad\quad\quad-\frac{d}{dz}H^{-1}(z)\delta\bar{z}\Big\}^{-1}  \Big\{\frac{dz}{d\tilde{z}}\Big\}^{-1} \nonumber\\
    &=&H(z)\Big\{1+3\delta\bar{z}-H^{-1}(z)\frac{d}{dz}H(z)\delta\bar{z}\Big\}  \Big\{\frac{dz}{d\tilde{z}}\Big\}^{-1}, \nonumber\\ \label{eq:C6}
\end{eqnarray}
where $H(z)$ is the unperturbed Hubble parameter and we have ignored the terms higher than first order. Taking derivative with respect to $\tilde{z}$ of $z=\tilde{z}+\delta z$ and taking account the identity of the conformal time $\eta$
\begin{eqnarray}d\eta&=&\frac{-d\tilde{z}}{H_0\sqrt{\Omega_m(1+z)^3+\Omega_r(1+z)^4+\Omega_k(1+z)^2+\Omega_{\Lambda}}}\nonumber\\&=&-H^{-1}(z)d\tilde{z},\nonumber\end{eqnarray}
we have
\begin{eqnarray}
    \frac{dz}{d\tilde{z}}=1+\delta\bar{z}-\frac{1+z}{H(z)}\Big(\dot{\Psi}-\textbf{n}\cdot\nabla\Psi\Big) \label{eq:C7}
\end{eqnarray}
The formulas above are model independent. Inserting Eq.~(\ref{eq:C7}) into Eq.~(\ref{eq:C6}), Hubble parameter $H(z,\textbf{n})$ reads
\begin{eqnarray}
    H(z,\textbf{n})
    &=&H(z)+2H(z)\delta\bar{z}-\frac{d}{dz}H(z)\delta\bar{z}\nonumber\\
    & &\quad\quad\quad\quad+\frac{1}{a(t)}\frac{\partial}{\partial \eta}\Psi-\frac{1}{a(t)}\textbf{n}\cdot\nabla \Psi \nonumber\\
    &=&H(z)+\frac{1}{2}H(z)\Psi-3H(z)\int^{\eta}_{\eta_0}{d\eta \dot{\Psi}}\nonumber\\
    & &\quad+\frac{3}{2}H(z)\textbf{v}\cdot\textbf{n}+\frac{1}{a(t)}\frac{\partial}{\partial \eta}\Psi-\frac{1}{a(t)}\textbf{n}\cdot\nabla \Psi. \nonumber\\
\end{eqnarray}
Here we consider the case of $\textrm{CDM}$ with
\begin{eqnarray}
    \frac{d}{dz}H(\textbf{n},z)=\frac{3}{2}(1+z)^{-1}H(\textbf{n},z)+\textrm{first order}. \nonumber
\end{eqnarray}
We could easily find that the perturbed Hubble parameter got from the perturbed luminosity distance is exactly the same with that got from the velocity field which can be seen in Eq.~(\ref{eq51}).

\section{SOME SPECIAL FUNCTIONS AND THEIR PROPERTIES}
Complex exponential function can be decomposed into spherical Bessel function and spherical harmonics function:
\begin{eqnarray}e^{i\textbf{k}\cdot\textbf{x}}=\sum_{\ell m}4\pi i^\ell j_\ell(kx)Y_{\ell m}(\hat{\textbf{k}})Y^*_{\ell m}(\textbf{n}) , \label{eq:B1} \end{eqnarray}
and
\begin{eqnarray}i\textbf{k}\cdot\textbf{n}e^{i\textbf{k}\cdot\textbf{x}}=\sum_{\ell m}4\pi i^\ell kj'_\ell(kx)Y_{\ell m}(\hat{\textbf{k}})Y^*_{\ell m}(\textbf{n}) ,\nonumber \end{eqnarray}
where $x=(\textbf{x}\cdot\textbf{x})^{1/2}$, and $\textbf{n}$ is the unit vector along direction $\textbf{x}$. $j_\ell(x)$ is the spherical Bessel function and satisfies
\begin{eqnarray}j'_\ell(x)&=&j_{\ell-1}-\frac{\ell+1}{x}j_\ell(x), \label{eq:B2}\end{eqnarray}
and
\begin{eqnarray}j_\ell(x)&=&\sqrt{\frac{\pi}{2x}}J_{\ell+1/2}(x). \label{eq:B3}\end{eqnarray}
$J_{\ell}(x)$ is the Bessel function and what we used most is the Schafgeitlin integral formula:
\begin{eqnarray}\int_0^\infty& &{\frac{J_\mu(ax)J_\nu(bx)}{x^\lambda}dx}=\frac{b^{\nu}\Gamma(\frac{\nu+\mu-\lambda+1}{2})}{2^\lambda a^{\nu-\lambda+1}
    \Gamma(\nu+1)\Gamma(\frac{\mu-\nu+\lambda+1}{2})}\nonumber \\
    & &\times F\Big(\frac{\nu+\mu-\lambda+1}{2},\frac{\nu-\mu-\lambda+1}{2};\nu+1;\frac{b^2}{a^2} \Big),\nonumber \\
   & &\qquad \mu+\nu+1>\lambda>-1,0<b<a\quad or  \nonumber \\ & &  \quad\quad \mu+\nu+1>\lambda>0, a=b .\label{eq:B4}\end{eqnarray}
Here $\Gamma(z)$ is the gamma function, and satisfies
\begin{eqnarray}\Gamma(z+1)=z\Gamma(z).\qquad   \label{eq:B5}\end{eqnarray}
$F(a,b;c;z)$ is the Gaussian hypergeometric function which follows
\begin{eqnarray}F(a,b& &;a-b+1;z)=(1+\sqrt{z})^{-2a}\nonumber \\
& &\times F\Big(a,a-b+\frac{1}{2};2a-2b+1;\frac{4\sqrt{z}}{(1+\sqrt{z})^2} \Big) \label{eq:B6}\end{eqnarray} and
\begin{eqnarray}\frac{d}{dz}F(a,b;a-b+1;z)=\frac{a\cdot b}{c}F(a+1,b+1;c+1;z). \nonumber \\ \label{eq:B7}\end{eqnarray}
$Y_{\ell m}(\textbf{n})$ in Eq.~(\ref{eq:B1}) is the spherical harmonics function with
\begin{eqnarray}\int{d\Omega_{\textbf{n}} Y^*_{\ell m}(\textbf{n}) Y_{\ell' m'}(\textbf{n})}=\delta_{\ell\ell'}\delta_{mm'}. \label{eq:B8}\end{eqnarray}
And the sum on $m$ along direction $\textbf{n}$ and $\textbf{n}'$ is
\begin{eqnarray}P_\ell(\textbf{n}\cdot\textbf{n}')=\frac{2\ell+1}{4\pi}\sum_{m} Y^*_{\ell m}(\textbf{n}) Y_{\ell m}(\textbf{n}'),
 \label{eq:B9} \\ \nonumber \end{eqnarray}
where $P_\ell(\textbf{n}\cdot\textbf{n}')$ is the Legendre function. Finally, it's notable that the indexes $\mu$, $\nu$, $\lambda$ and $\ell$ are all real numbers \cite{b34,b35}.


\begin{thebibliography}{99}
\bibitem{note}
The terms analytical result and numerical result appear frequently. The analytical result represents the result from concrete expresion or identity appeared in books or references. While the numerical result means the result obtained by numerical computation, for example, in this paper double Bessel function is integrated by Newton-Cotes method.
\end{thebibliography}
\end{document}